# Implications for Governance in Public Perceptions of Societal-scale AI Risks

Ross Gruetzemacher[1,2,3,]*, Toby D. Pilditch[2,4,5], Huigang Liang[6], Christy Manning[1,2], Vael Gates[7], David Moss[8], James W. B. Elsey[8], Willem W. A. Sleegers[8], and Kyle Kilian[2,9]

[1]Wichita State University, [2]Transformative Futures Institute, [3]Centre for the Study of Existential Risk, University of Cambridge, [4]University of Oxford, [5]University College London, [6]University of Memphis, [7]Arkose, [8]Rethink Priorities, and [9]Center for the Future Mind, Florida Atlantic University

## Abstract:

Amid growing concerns over AI's societal risks—ranging from civilizational collapse to misinformation and systemic bias—this study explores the perceptions of AI experts and the general US registered voters on the likelihood and impact of 18 specific AI risks, alongside their policy preferences for managing these risks. While both groups favor international oversight over national or corporate governance, our survey reveals a discrepancy: voters perceive AI risks as both more likely and more impactful than experts, and also advocate for slower AI development. Specifically, our findings indicate that policy interventions may best assuage collective concerns if they attempt to more carefully balance mitigation efforts across all classes of societal-scale risks, effectively nullifying the near-vs-long-term debate over AI risks. More broadly, our results will serve not only to enable more substantive policy discussions for preventing and mitigating AI risks, but also to underscore the challenge of consensus building for effective policy implementation.

## One Sentence Summary:

US voters and experts agree that international bodies should govern AI risks but voters consider the societal-scale risks of advanced AI as more likely and more impactful than experts, preferring a slower pace of development.

## Background

Worries about societal-scale risks from powerful artificial intelligence (AI) systems are nearly as old as the field itself [1], and as capabilities of systems have grown in recent decades so has research on policy, governance, and technical topics seeking to ensure that the development of AI is safe and beneficial to humanity [2]. Over the past decade, governments have been increasingly aware of the potentially profound significance of AI [3], and since the release of ChatGPT in November of 2022, the potential of AI has become obvious to the broader public.

The reason for this interest is not unwarranted. As a general purpose technology, AI's potential to do great good or great harm is unparalleled [4]. AI-generated advances could cure diseases [5], foster a new economic age [6], and help to tackle global challenges like climate change [7].

Conversely, the threats AI could pose are both wide-ranging and stark. While existential risks like extinction, civilizational collapse, or dystopian futures grab immediate attention, other societal-scale risks such as algorithmic bias, knowledge deterioration, and economic concentration of power have tarted to emerge and be recognized. [6,8-9]. Moreover, AI technology both changes and is changed by the human behaviors and systems surrounding it [10] yielding risks that stem from structural interactions (e.g., growing geopolitical instability, erosion of trust).

It is not surprising that governments have already begun efforts to regulate AI development (e.g., US Executive Order 14110), a topic with which they are substantively unfamiliar. Consider that in 1996 the FCC was faced with the challenge of determining suitable regulation for the (new) internet. In so doing, they had to balance prospective economic and societal gains with potential risks, both known and unknown. The shape this watershed legislation took, and the impact it had on fostering the nascent technology in a consumer-protected, but innovation-enabling manner, shaped not only the standards and expectations of this new information age, but the way society and the economy developed in the decades that followed [11].

In the advent of new and powerful technologies, regulatory policies are shaped by an emergent balance between industry stakeholders, experts, governments, and voters/consumers that drives both democratic and economic forces. Although all parties have their respective incentives, divergence of opinion can be especially consequential (e.g., public vs expert opinion on risks of nuclear power [12]). Therefore it is important for governments to thoroughly understand the varying perspectives of these different parties, not to rely solely on the opinions of subject matter experts, but to also consider perspectives of the citizen-consumers who hold collective power.

**Survey**

To address this knowledge gap, we administered a survey of the AI risk perceptions and policy preferences of 120 AI experts and 400 US registered voters. The survey covered perceptions of both the likelihood and prospective impact of 18 specific AI risk scenarios. These scenarios comprise seven classes of societal-scale risks from advanced AI: economic, ethical, misuse, accident, geopolitical, environmental, and existential. Generally, the societal-scale AI risks described for the 18 scenarios comprising these seven classes concern risks of AI-induced harms to large-scale social systems (e.g., financial systems, geopolitical stability) or to nations or other large social groups if the outcomes (e.g., human rights violations, economic harms, war) of these harms are sufficiently widespread, and are inclusive of catastrophic and existential risks. We emphasize that we do not draw a distinction between short and longer-term risk (e.g., [13]), but rather classify the broad areas from which societal-scale harms may arise.

Additionally, to understand perceptions of different approaches to governing AI to mitigate risks, we asked respondents two questions on perceived regulatory best practices for risk mitigation. These questions concerned the speed with which AI progress should proceed—should it be paused, slowed down, maintained, or accelerated—and who should bear the responsibility for managing AI risk: companies, governments, or international institutions.

In contrast to previous work [14-15], this study is the first survey to conduct a comparison on both the likelihood and impact of risks across the breadth of all societal-scale AI harms—from bias/discrimination concerns, privacy issues, and economic concerns to terrorist AI weapons, an AI arms race, and existential concerns. Moreover, unlike the previous work, it is not limited to either experts or the general public but compares the results from both groups, focusing specifically on US voters given the relevance of their opinions to US policymakers.

## Perceived Societal-Scale Risk Likelihood vs Impact

Through the disentangling of risk likelihood and risk impact, we find that US registered voters (Fig. 1A) estimate not only the likelihood of societal-scale risks stemming from AI as significantly higher than AI experts (Fig. 1B), $\beta = 0.306$, SE = 0.040, t = 7.748, $p < 0.001$, they also estimate the prospective impact of those risks as significantly higher, $\beta = 0.649$, SE = 0.033, t = 19.47, $p < 0.001$. Although this exposes potential tensions in the incentives for AI regulation, we discover a similar pattern across groups regarding the *relative* likelihoods and impacts of risks. For instance, while both groups deem existential categories of risk (e.g., civilizational collapse) as substantially less likely than others, including economic (e.g., economic instability) or ethical (e.g., privacy) risk categories, $\beta = -1.489$, SE = 0.099, t = -15.10, $p < 0.001$, the former is considered substantially more impactful, $\beta = 0.958$, SE = 0.086, t = 11.09, $p < 0.001$.

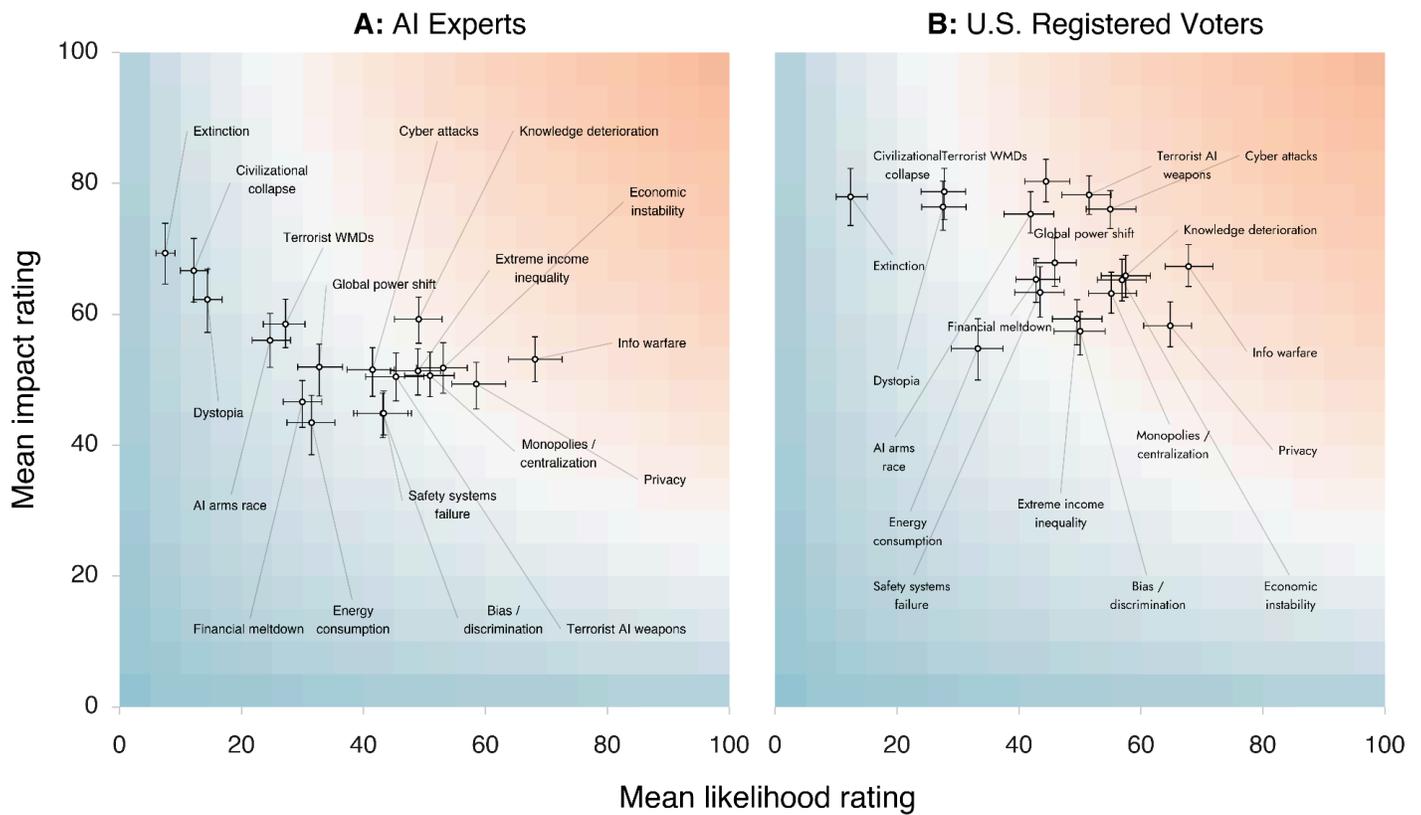

**Fig. 1. Perceptions of societal-scale AI risks depicted as risk matrices for 18 scenarios in seven classes of societal-scale AI risks.** Fig. 1A (left): Experts' perceptions of societal-scale AI risks. Fig. 1B (Right): US registered voters' perceptions of societal-scale AI risks depicted identically. Error bars reflect 95% Highest Density Intervals (HDIs).

Although this broad agreement is useful for considerations of relative risk/safety prioritization, we note that US voters deem the potential impact of misuse societal-scale risks (e.g., terrorist AI weapons) as significantly more damaging (relative to other risks) in comparison to AI experts, $\beta = 0.595$, SE = 0.11, t = 5.19, $p < 0.001$, and more likely, $\beta = 0.319$, SE = 0.145, t = 2.205, $p < 0.05$. In addition, US voters deem existential risks as more likely than AI experts, $\beta = 0.656$, SE = 0.171, t = 3.839, $p < 0.001$. Whether this divergence represents underestimation/complacency/optimism among AI experts (e.g. due to structural incentives to dismiss or minimize risks the experts themselves may be causing), or overestimation/alarmism/pessimism among voters (e.g., due to salient, emotive topic areas like terrorism) is indeterminate.

To further clarify the nature of voter and expert associations between various AI risk likelihoods and impacts, we performed a correlation analysis to determine potential common and distinct clusters (defined as positive correlations > 0.5; full details in the supplementary materials). In so doing, four distinct clusters of risk impacts (Fig. 2A) are shared by both experts and voters (solid lines): misuse (e.g., terrorist WMD attacks), catastrophic risks (e.g., civilizational collapse), information (e.g., information warfare and ecosystem collapse), and geopolitical threats (e.g., authoritarianism and great power war). With the exception of the latter, we see the same common clusters (solid lines) associated by both voters and experts for risk likelihoods (Fig. 2B), indicating strong general consistency of associations among both groups on these areas of AI risk.

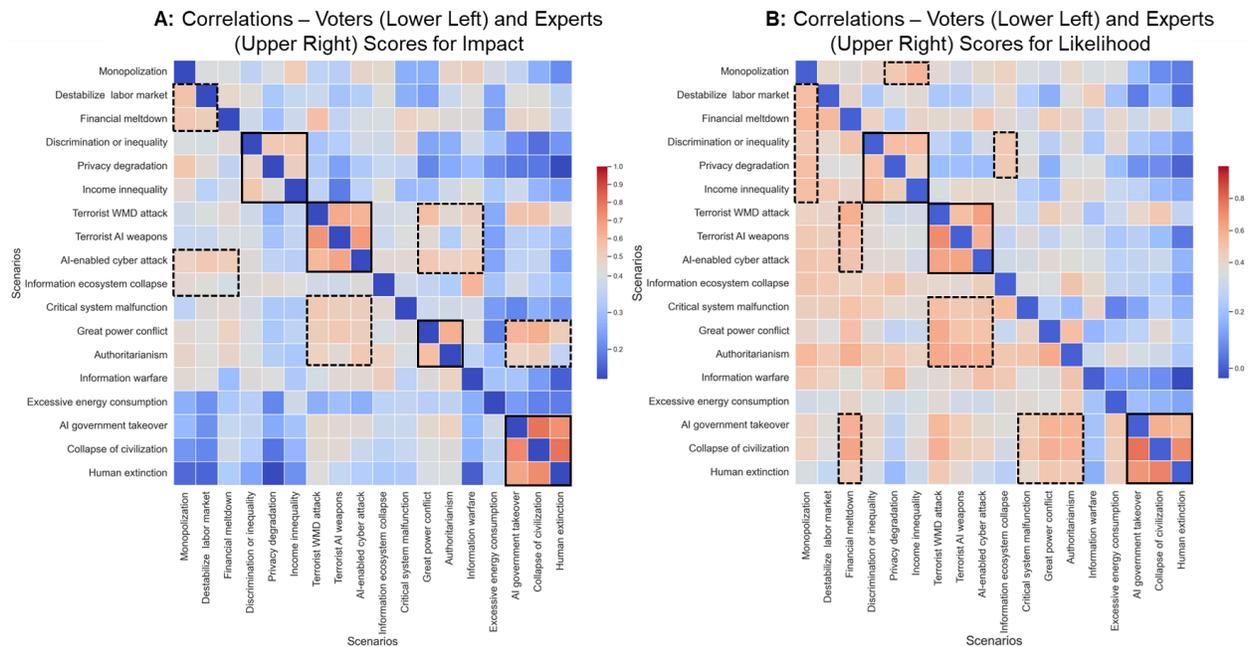

**Fig. 2. Thematic clusters and their correlations between experts and voters.** A correlation analysis between likelihood (**2A**) and impact (**2B**) scores for voters (lower left quadrant) and experts (upper right quadrant) displays distinct clusters where voters' and experts' responses trend together (**black line**,

correlation > 0.5) or separately (**dashed line**, correlations > 0.5), highlighting which themes were weighted similarly or differently by voters and expert participants.

However, distinct clusters (dashed lines) corroborate the voter-expert differences outlined above. Specifically, whilst experts show a unique cluster between the impacts (Fig. 2A) of great power conflicts and authoritarianism with catastrophic risks, voters show a unique cluster of AI cyberattacks and information ecosystem collapse with economic impacts (e.g., financial meltdown), indicative of voters greater concern for these more immediate* risks. In considering associations of AI risk likelihoods (Fig. 2B), we find few distinct associations of likelihoods among experts, whilst voters have several distinct clusters of note, such as the clustering of socio-economic risks and monopolization, misuse and geopolitical risks, geopolitical risks and catastrophic risks, and finally catastrophic risks (and separately, misuse risks) and financial meltdown. This suggests that whilst experts typically consider these latter risk likelihoods more independent, voters commonly link misuse, to political and economic risks, and those in turn to catastrophic risks.

## Policy Preferences

Importantly, we find broad agreement across AI experts and US registered voters regarding who should be responsible for managing AI risk (Fig. 3A). While a plurality of both US voters and AI experts prefer international treaties, intergovernmental organizations, and NGOs to manage risks from advanced AI, fewer voters and experts believe either tech companies or national governments should be responsible for managing AI risks, with US voters being significantly more wary than AI experts of national governments in this regard, $\chi^2(3) = 17.92$, $p < .001$. This may reflect recognition among both groups of the potentially flawed incentives and/or capability of the latter groups to effectively (self)regulate.

However, despite agreeing on who should manage risk, we find US registered voters and AI experts disagree significantly on what should be done about the pace of AI development (Fig. 3B). While experts believe the pace of AI development should be accelerated or maintained more than voters, voters preferred to slow down the pace of AI development as compared to experts, $\chi^2(3) = 56.04$, $p < .001$. This is in broad alignment with the globally higher perceptions of both risk likelihoods and impacts from AI among US voters.

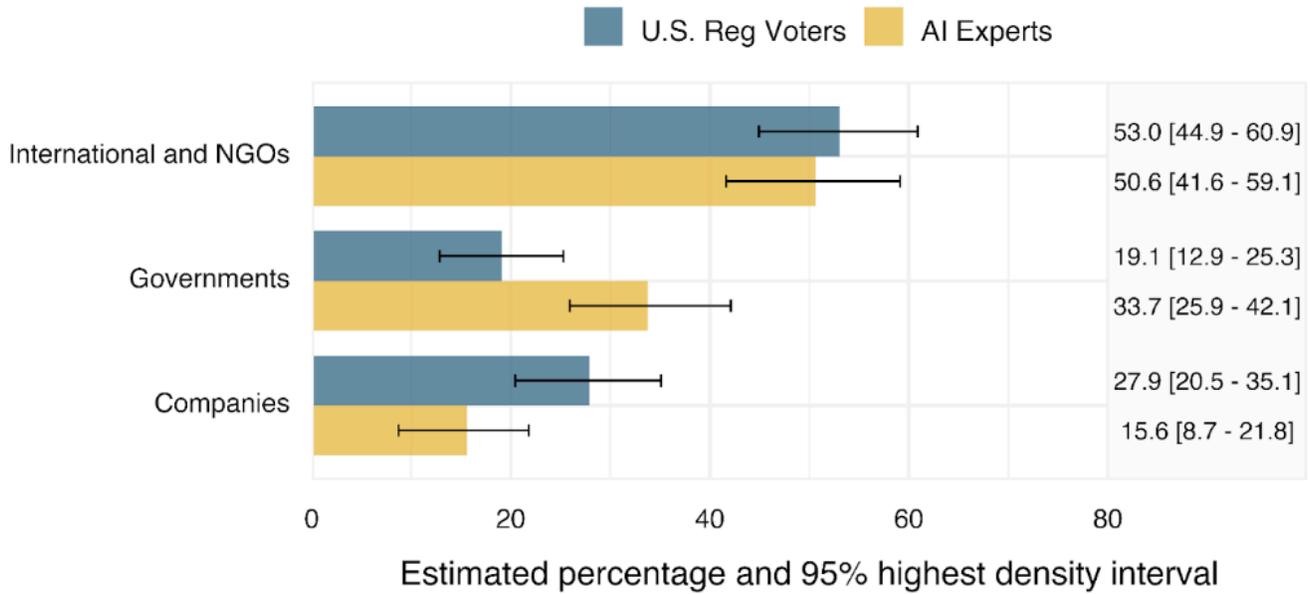

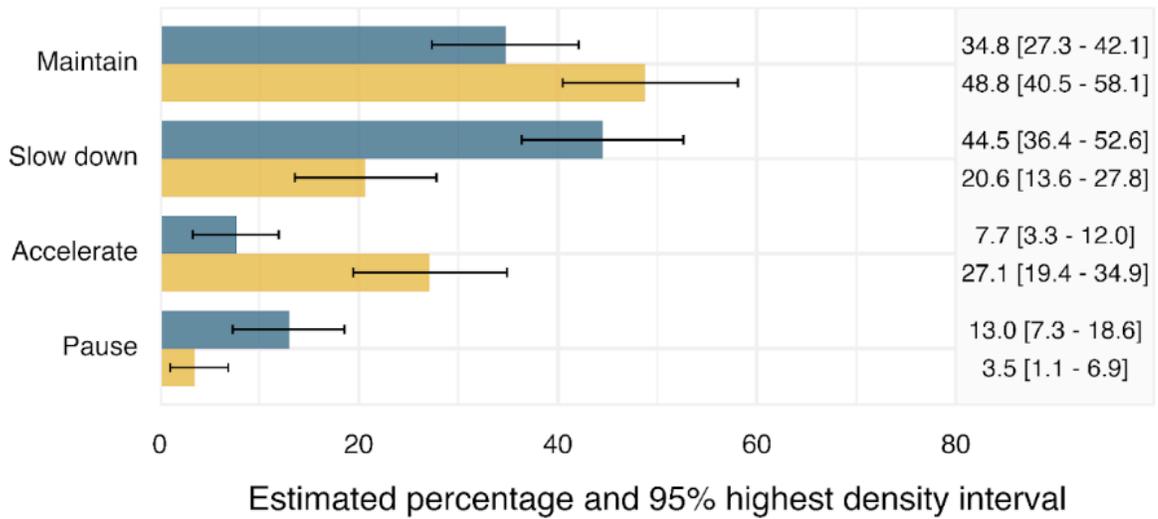

**Fig. 3. Perspectives on the management of AI risks and the preferred pace of AI progress by U.S. registered voters and AI experts.** Fig. 3A. (Top): Respondents preferences for who should manage risks from advanced AI, split by group. Fig. 3B. (Bottom): Preferences for pace of advanced AI development, split by group. Error bars reflect 95% HDIs.

# Discussion

The question of how to mitigate societal-scale risks arising from the development of advanced AI is one that is increasingly relevant and not easily solved. Both the positive and negative outcomes associated with the development of ever more capable AI systems stand to affect us all in ways we cannot fully anticipate [4]. Given the shared stakes and the substantial uncertainties involving the management of advanced AI development, it is valuable not only to understand experts' perspectives on AI risks and their mitigation, but also those of the voting public, as each have important roles to play in informing governments' regulatory and strategic decisions.

Our findings shed new light on the perceptions of societal-scale risks among both AI experts and US registered voters. We find that US voters generally perceive the societal risks associated with AI as both more likely and more potentially damaging (i.e., impactful) than experts, and notably in regard to the potential misuse of AI tools (e.g., terrorist actions). More broadly, we find a consistent characterization of risk classes across both groups, demonstrating distinct clusters for medium likelihood, medium impact risks (e.g., economic and accident risk classes), and distinct clusters for low likelihood, high impact risks (e.g., existential). Further, our supplementary materials include figures containing heatmaps on each risk scenario for both voters and experts, enabling more nuanced comparisons.

By disentangling the perceived likelihood of risks from their perceived impact, we can more clearly characterize commonalities and avoid misunderstandings when determining risk area priorities for both research and policy, and conversely highlight areas of disagreement more efficiently. For example, a disagreement about risk importance where the difference stems from perceived likelihood indicates a need to better understand and agree the likely *causes* and predicates of a risk. Conversely, a disagreement stemming from impact implies a need to focus on the chain of events and mechanisms of *effects* incurred by the risk. Further, in mapping these risk class profiles, we lay the foundation for more effective policy design. For example, where risks are characterized by high likelihoods, policy goals can center around risk likelihood reduction (i.e., prevention) strategies, while those characterized by high impact and low likelihoods are inherently more challenging and beyond the scope of traditional wicked problems.

These findings also appear to have implications for the near-vs-long-term debate over how to prioritize mitigation of different risks from advanced AI [13]. Rather than finding in favor of prioritizing nearer-term risks [16] over the more extreme longer-term risks [17], they demonstrate that experts and voters both share more nuanced perceptions of societal-scale AI risks. Consequently, concerns over such risks might be best assuaged through a more nuanced approach which would require more carefully balancing mitigation efforts among all classes of societal-scale risks. Moreover, sociotechnical AI risks [10, 18] appear to be considered as similarly impactful and probable for both experts and voters as risks from accident and misuse [8], and as such should not be marginalized in overall risk mitigating policy planning [19].

Our policy preference findings reveal a consistent belief that risk management should be the task of international treaties, intergovernmental organizations and NGOs (>50% of voters and experts), while few trust tech companies with such a responsibility: ~26% voters and ~15% experts. In this regard, and consistent with our risk perception findings, we note that voters are more wary than experts of the current pace of AI development, with significantly higher proportions believing it should be slowed or paused. These findings also suggest that voters trust governments less than experts in managing AI risk, suggesting that legislators could benefit from public trust-building measures and outreach on this important topic.

Governments, including in the US, are actively working to develop legislation to address the risks posed by advanced AI development. Therefore, democratic governments should be informed about the perceptions of their constituents and those of experts in the field to make informed policy decisions. We note that these perceptions are not static, as opinions have no doubt shifted in the past year alone. However, just as there are important responsibilities to educate the public regarding these technologies, democracies are subject to the whims of voters and thus may be limited in their ability to respond to rapid technological progress. Having a robust and accurate understanding of both experts and voters is crucial if we are to effectively reach agreement and alignment in the governing of the safe, secure, and trustworthy development of AI.

## Acknowledgments

**Funding:** Funding for the study was provided by the Transformative Futures Institute, a nonprofit research organization based in Wichita, Kansas, United States.

**Human subjects:** This study was approved by Wichita State University's Institutional Review Board, Protocol ID: #5538, Protocol Title: Assessing Experts' Opinions on Societal-scale Risks from Artificial Intelligence (Expedited). All participants acknowledged an approved Institutional Review Board information letter demonstrating informed consent prior to beginning the survey.

**Author contributions:** Conceptualization: RG, VG; Methodology: RG, HL, TP, CM, VG, KK; Formal analysis: JE, TP, WS; Data curation: WS, JE; Investigation: DM; Visualization: JE, WS, TP, KK; Project administration: RG; Supervision: RG, DM; Writing—original draft: TP, KK, RG; Writing—review & editing: TP, KK, RG, CM, VG, HL.

**Data and materials availability:** Anonymized data from this study can be found at www.github.com/rossgritz/societal-scale-ai-risks.

# Supplementary Materials

## Methods

We administered our survey to 120 AI experts and 400 US registered voters. The survey was conducted within the qualtrics platform. All participants provided informed consent prior to participation and the study in accordance with the IRB protocol.

*Participants*

AI experts were qualified as being authors of papers at one of the top three machine learning conferences (i.e., the International Conference on Machine Learning, the International Conference on Learning Representations, the Annual Conference Neural Information Processing Systems) over the past three years. 8,000 experts were contacted via email (see Materials B) with the offer of a $50 gift card compensation for 10 minutes of their time. Of these, 120 experts consented to participate, with 118 completing the study. 99 experts identified as male, 13 as female, and 6 preferred not to say. Experts were recruited globally, and aged between 23 and 54 years old (*Mean* = 31.9; *SD* = 6.71 years), with 107 self-identifying as AI experts (probably or definitely an expert). The median time for completing the survey was 10.28 minutes, resulting in an effective hourly wage of $292/hr.

US voters were recruited via the CloudResearch platform Connect (https://www.cloudresearch.com/), with the study advertised as a "Survey About Societal Issues" (full advertisement depicted in Materials C, below), and offering $10 compensation for participation. The minimum age requirement was 18 years old. 400 participants were recruited for the study, from which 4 were removed for failing all three attention checks, resulting in a voter sample size of 396. Of these participants 196 identified as male, 182 as female, 2 as non-binary or other, and 1 preferred not to say. Voters were aged between 19 and 85 years old (*Mean* = 44.1; *SD* = 14.39 years), and 217 held at least a Bachelor's degree. The median time for completing the survey was 10.3 minutes, resulting in an effective hourly wage of $58.25/hr.

All participants were required to provide informed consent after reading the approved consent form. This was in accordance with the Wichita State University Institutional Review Board's approved Protocol #5538 for this study, titled "Assessing Experts' Opinions on Societal-scale Risks from Artificial Intelligence". Originally the study was to focus on experts, and a modification form was submitted and approved prior to the start of the study to include US registered voters in order to enable a comparison.

*Procedure*

Recruitment of both experts and voters took place in October 2023, over a three week period. Experts were recruited via an email circular, while US voters were recruited via CloudResearch's Connect platform (Hartman et al. 2023). Following recruitment, all participants first provided informed consent and then proceeded with the survey.

The survey (see Materials A below) consisted of paired questions regarding the perceived likelihood and perceived impact on each of 18 societal-scale risks associated with advanced AI. Responses were given on a labeled slider (0-100% with 20% markers for likelihoods, and evenly spaced labels of Negligible, Minor, Moderate, Major, Catastrophic for impact), and the 18 risks were presented in a randomized order, with each presented on a separate page of the survey. Following this, participants answered two questions regarding the management of risk: who should manage risks from advanced AI? (forced choice: tech companies, national governments, or international treaties, organizations and NGOs), and what should we do about risks from advanced AI? (forced choice: pause development, slow down development, maintain development, accelerate development). Finally, all participants completed a demographics questionnaire covering basic demographics, as well as their expertise in AI, current profession, and educational background (see Materials A below). Upon completion of the survey, all participants were debriefed and compensation provided.

*Analysis*

Data for U.S. Registered voters was weighted to account for Age (18-24, 25-44, 45-64, 65+), Racial identity (Black or African American, Hispanic or Latino, White or Caucasian, Other), Gender (Man, Woman), Education (High school or less, Some college no degree, Graduated from college, Completed graduate school), Political Party Identification (Republican, Democrat, Independent/Other/Don't know), Family income (Under $20k, $20k-$49k, $50-$79k, $80k-$99, $100-$149k, At or above $150k), Pew 'Born again' Christian status (Yes, No), reported Urbanity/Rurality (City, Rural Area, Suburb, Town), US Census Region (Midwest, Northeast, South, West), and 2020 Presidential vote (Biden, Trump, Did not vote but eligible to vote). Target proportions for these variables were generated using the registered voter weights of the Cooperative Election Study (Schaffner et al. 2022). Weights were generated using the ANES raking algorithm present in the anesrake (Pasek & Pasek, 2018) R package.

Weighted analyses were conducted using weighted Bayesian regression models in brms (Burkner, 2017), with sample size penalized proportional to the observed design effect from the weighting procedure (in this case, 2.8) to produce accurate (and wider) uncertainty intervals around point estimates. For Importance and Likelihood estimates, a beta regression model was used, with the mean outcome rating and precision predicted by the type of risk, and observations nested within subjects. Categorical regression was used for the 'Speed' and 'Management' outcomes, in this case with only the Intercept included in the model. The beta regression model used weakly informative priors for the Intercept (normal(0, 1)), effect of each risk (normal(0, 1)), standard deviation of the intercept by respondent (exponential(2)), and precision (exponential(1), with a lower bound of .001). A weakly informative prior was also placed on the intercept for the categorical model (normal(0, 1.5)).

For the participant correlation study, we developed a methodological approach to visually compare the perceptions of both experts and voters on societal-scale risk scenarios through the use of a combined correlation matrix heatmap, which distinctly allocated voters' ratings to the lower triangle and experts' ratings to the upper triangle for two distinct impact and likelihood heatmaps. This was done through systematic data preparation that involved the imputation of missing values using column-wise means followed by the calculation of Pearson correlation coefficients. Each correlation matrix was bifurcated into lower and upper triangles, representing voters and experts, respectively. The final matrices were visualized using the Python Seaborn library centered around the mean correlation value to accentuate deviations in perceptions between the two groups, facilitating an intuitive comparative analysis of alignment or discordance in their assessments.

For impact correlations, the specific correlation values were: terrorist WMD attacks showed strong positive correlations with terrorist AI weapons (0.66 for experts and 0.71 for voters) and AI-enabled cyber attacks (0.63 for experts and 0.6 for voters), and between AI government takeover, civilizational collapse (0.81 for experts and 0.74 for voters), and human extinction (0.73 for experts and 0.66 for voters). Information warfare and information ecosystem collapse were positively correlated for both expert and voter responses.

*Additional Results*

Below, in Fig. S1 and Fig. S2, we provide additional information detailing the results presented in the main text. Further below, in Fig. S3 and Fig. S4, we provide further additional information supporting discussion of experts vs. voters on extreme risks.

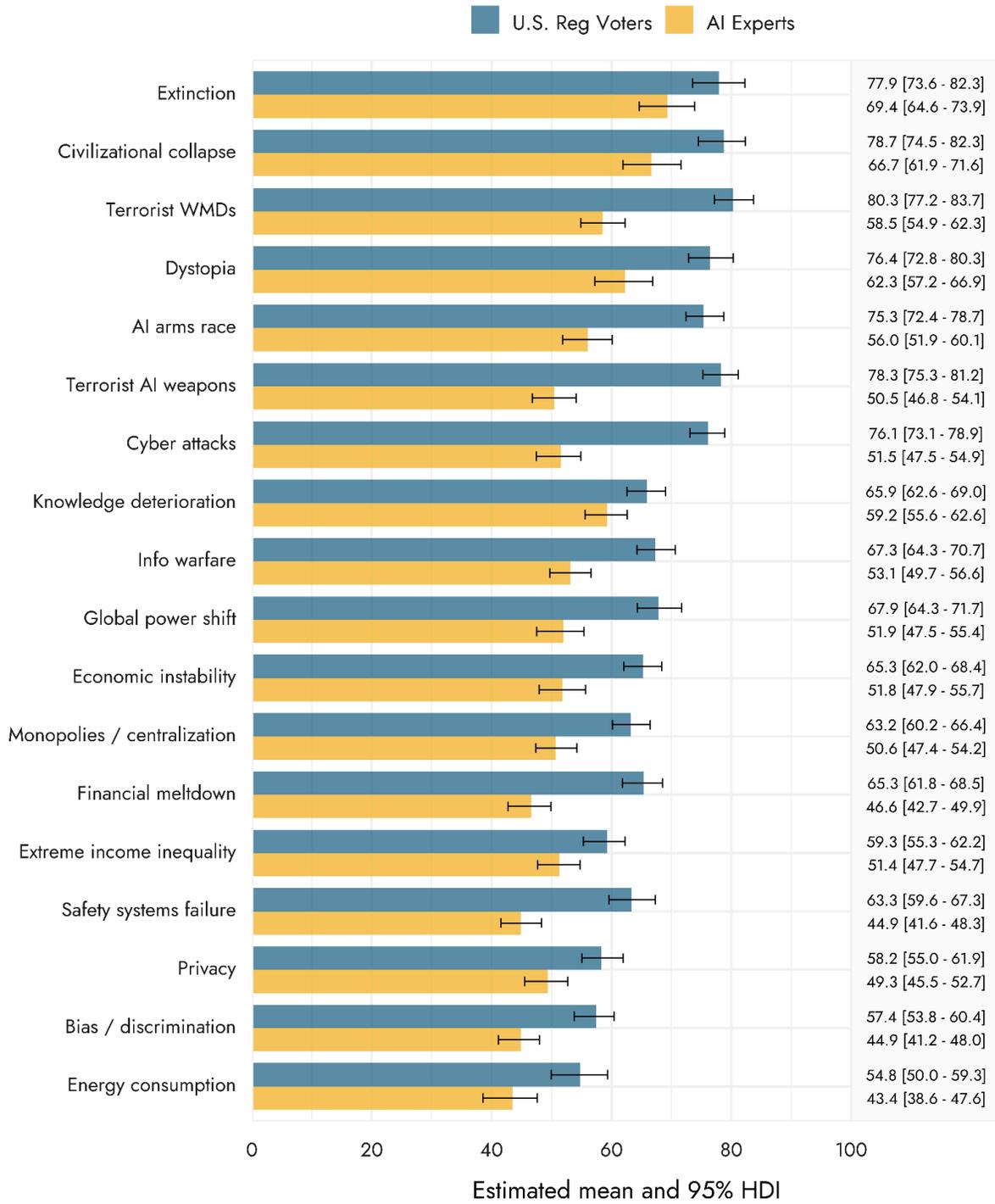

**Fig. S1 : Comparison of Impact Ratings Between Experts and Voters.**

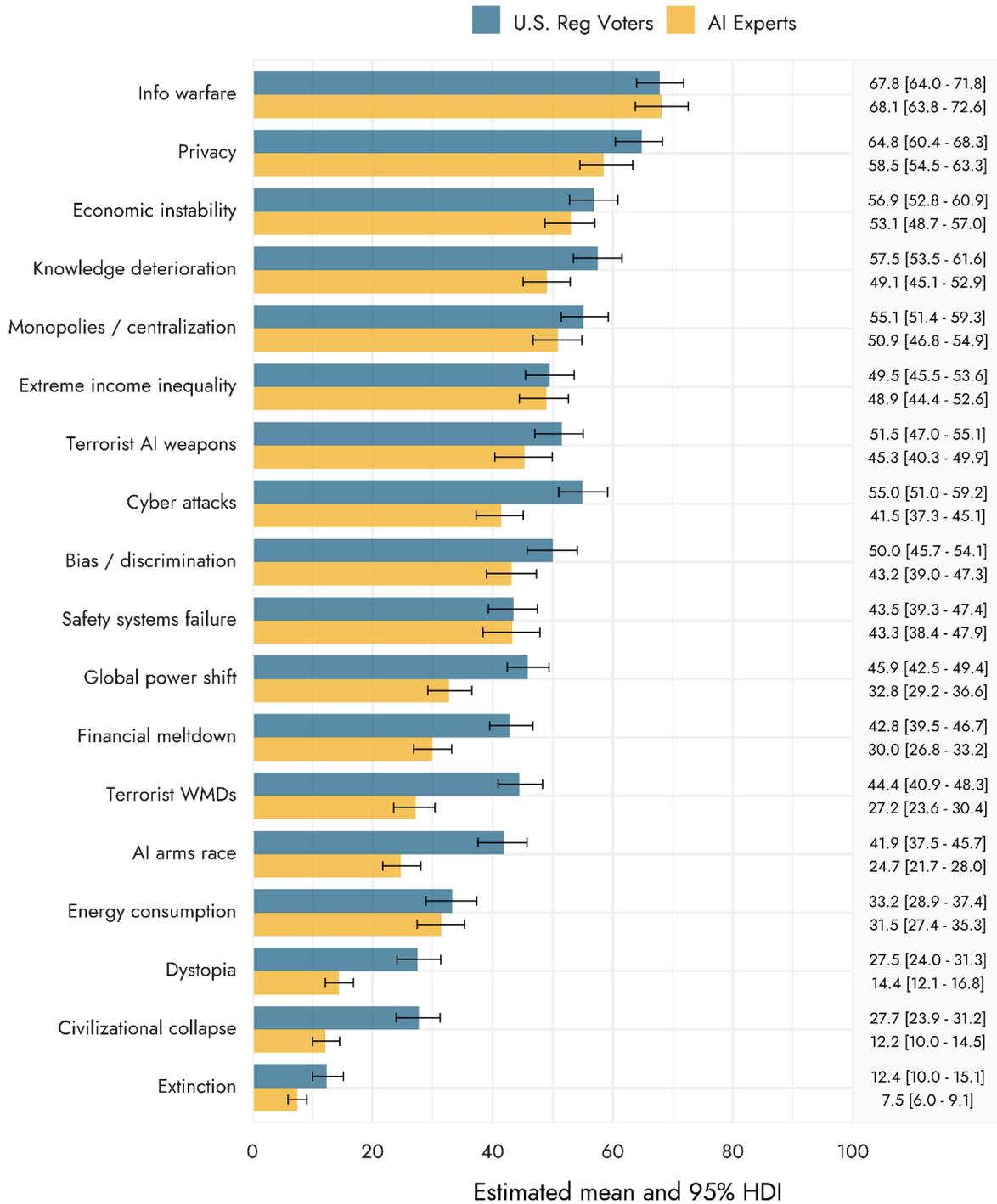

Fig. S2: Comparison of Likelihood Ratings Between Experts and Voters.

Below, Fig. S3 and Fig. S4, we provide further additional information detailing the results presented in the main text.

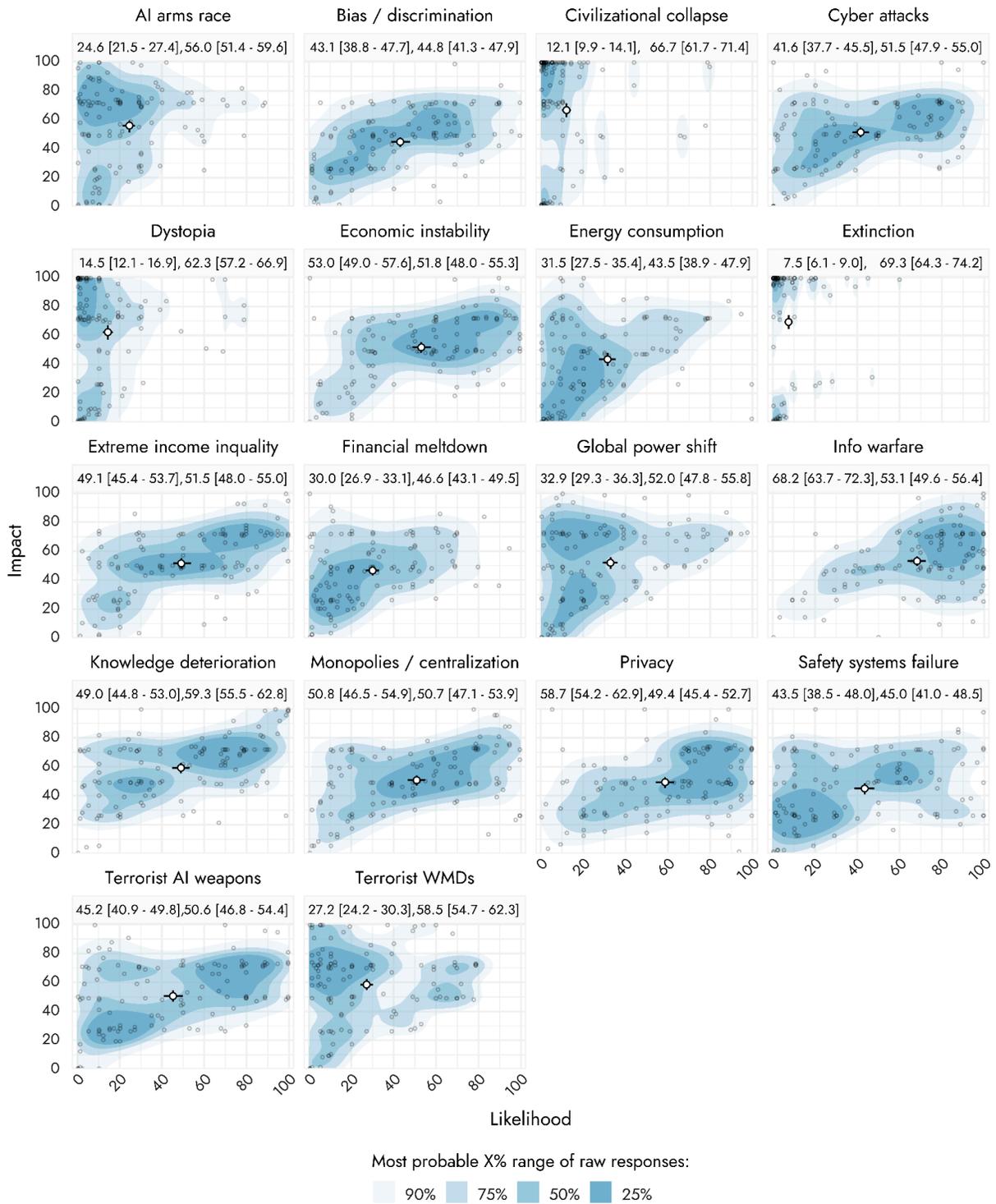

**Fig. S3: Heatmaps for experts' likelihood and impact ratings on each question.**

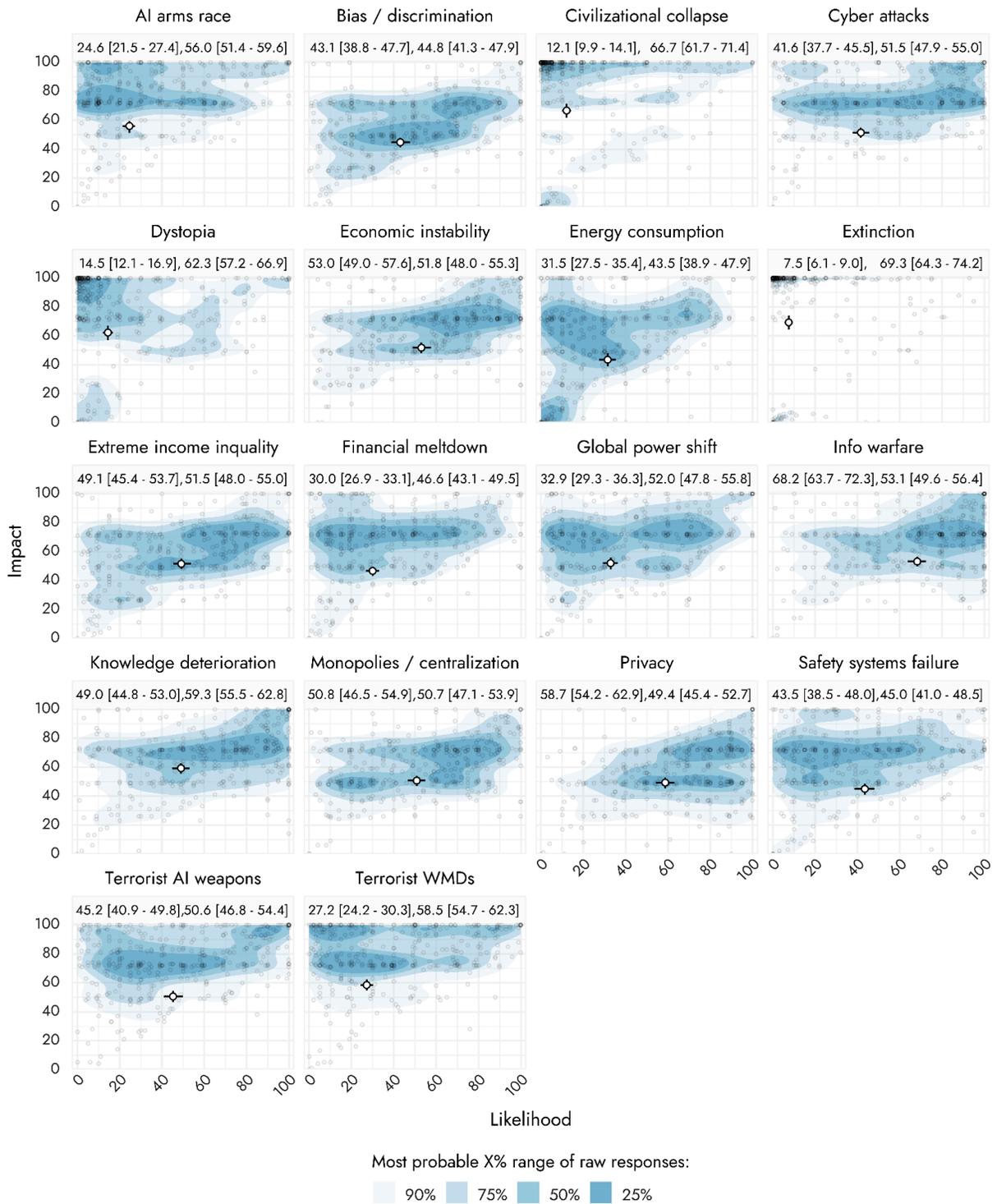

Fig. S4: Heatmaps for voters' likelihood and impact ratings on each question.

*Limitations*

A key limitation for the study is representativeness of the different samples. For the experts, we do not have information about the total population of AI experts, and so it is not possible to determine the extent to which the expert sample accurately reflects the views of AI experts more generally; selection biases could result in especially concerned or unconcerned experts taking this survey, for example. For the U.S. registered voter population, while the sample was weighted to be representative, the sample size was relatively small for this purpose, and it is therefore possible that some views are underrepresented.

## Materials A: Societal-scale AI Risks Survey

The following survey contains questions on 18 different scenarios of societal-scale risks from *advanced artificial intelligence (AI)*. We use the term 'advanced AI' to describe next-generation AI technology that is much more capable than current AI technology; this includes next-generation generative AI technology or robotics. In the coming decades, advanced AI is expected to transform society as it outperforms humans at an increasing number of tasks. This survey is trying to understand what risks from this technology are most concerning.

For each of the 18 scenarios, you will be asked about the likelihood and the impact of the scenario.

When completing the questions, please consider whether each scenario being discussed could occur within a decade or two.

At the end of the survey, you will be asked one final question as well as some brief demographic questions.

*Economic Risks*

**1. Monopolies & Centralization of Power**
   A. What is the likelihood that in the next decade or two advanced AI will lead to monopolies or concentration of power in a small number of companies? For example, a small number of companies might become so valuable from advanced AI that they have extreme influence on markets, the economy, or the government.

   [*Slider from 0% to 100%; labels every 20%, with 1% increments; default at 0*]

   B. How impactful do you think that monopolies or the centralization of economic power from advanced AI might be for society?

[*100 pt Slider with labels Negligible, Minor, Moderate, Major, and Catastrophic evenly spaced; default at 0 (Negligible)*]

**2. Economic Instability**

A. What is the likelihood that in the next decade or two advanced AI causes a major disruption of the economy, destabilizing labor markets? For example, advanced AI might make many skills obsolete, forcing some people to learn new skills to keep their jobs, and resulting in significant job losses for others.

[*Slider from 0% to 100%; labels every 20%, with 1% increments; default at 0*]

B. How impactful do you think the disruption of the economy by advanced AI might be for society?

[*100 pt Slider with labels Negligible, Minor, Moderate, Major, and Catastrophic evenly spaced; default at 0 (Negligible)*]

**3. Financial Meltdown**

A. What is the likelihood that in the next decade or two advanced AI leads to a global financial meltdown? For example, this might mean that automated decision making systems, such as in algorithmic trading systems with too much power and insufficient oversight, could result in a stock market crash.

[*Slider from 0% to 100%; labels every 20%, with 1% increments; default at 0*]

B. How impactful do you think financial problems caused by advanced AI might be for society?

[*100 pt Slider with labels Negligible, Minor, Moderate, Major, and Catastrophic evenly spaced; default at 0 (Negligible)*]

*Ethical Risks*

**4. Bias & Discrimination**

A. What is the likelihood that in the next decade or two biases embedded in advanced AI tools might lead to significant increases in discrimination and social or income inequality? For example, biases in advanced AI might increase disparities in hiring and wage determination, credit scoring, mortgage and rental application decisions, assessment of job performance, criminal sentencing and policing, school admissions decisions and personalized learning systems, or decisions in healthcare diagnosis and treatment.

[*Slider from 0% to 100%; labels every 20%, with 1% increments; default at 0*]

B. How impactful do you think increases in social or income inequality from advanced AI might be for society?

[*100 pt Slider with labels Negligible, Minor, Moderate, Major, and Catastrophic evenly spaced; default at 0 (Negligible)*]

### 5. Privacy

A. What is the likelihood that in the next decade or two advanced AI technologies cause a decrease in personal privacy protections? For example, advanced AI might make surveillance more common and more acceptable for preventing crime, or, it might increase the demand for value of training data for AI models tremendously, resulting in a loosening of privacy protections.

[*Slider from 0% to 100%; labels every 20%, with 1% increments; default at 0*]

B. How impactful do you think risks to privacy protections from advanced AI might be for society?

[*100 pt Slider with labels Negligible, Minor, Moderate, Major, and Catastrophic evenly spaced; default at 0 (Negligible)*]

### 6. Extreme Income Inequality

A. What is the likelihood that in the next decade or two advanced AI results in extreme income inequality? For example, this might happen if advanced AI drives productivity gains generate tremendous new wealth overwhelmingly benefiting those who are already very wealthy, while those with few or no investments see little or no benefits.

[*Slider from 0% to 100%; labels every 20%, with 1% increments; default at 0*]

B. How impactful do you think severe income inequality from advanced AI might be for society?

[*100 pt Slider with labels Negligible, Minor, Moderate, Major, and Catastrophic evenly spaced; default at 0 (Negligible)*]

*Misuse Risks*

### 7. Terrorist WMD Attack

A. What is the likelihood that in the next decade or two terrorists use advanced AI to create a deadly biological weapon or chemical weapon that is used at least once against a large number of civilians in a major terrorist attack?

[*Slider from 0% to 100%; labels every 20%, with 1% increments; default at 0*]

B. How impactful do you think terrorist attacks with deadly biological or chemical weapons developed with advanced AI might be for society?

[*100 pt Slider with labels Negligible, Minor, Moderate, Major, and Catastrophic evenly spaced; default at 0 (Negligible)*]

**8. Terrorist AI Weapons Attack**

A. What is the likelihood that in the next decade or two terrorists use advanced AI to create autonomous AI weapons for use in attacking civilians? For example, terrorists might use advanced AI to evade authorities and create cyberweapons and autonomous drones that could carry out attacks on critical infrastructure, such as hydroelectric dams or nuclear power plants, or on large civilian targets like airports or sporting arenas.

[*Slider from 0% to 100%; labels every 20%, with 1% increments; default at 0*]

B. How impactful do you think terrorist attacks from advanced AI weapons might be for society?

[*100 pt Slider with labels Negligible, Minor, Moderate, Major, and Catastrophic evenly spaced; default at 0 (Negligible)*]

**9. Cyber Attacks**

A. What is the likelihood in the next decade or two there are major cyber attacks on civilian targets enabled by advanced AI? For example, advanced AI might infiltrate and reprogram computer systems, like power grids or military computer systems, to execute destructive plans.

[*Slider from 0% to 100%; labels every 20%, with 1% increments; default at 0*]

B. How impactful do you think cyber attacks from advanced AI might be for society?

[*100 pt Slider with labels Negligible, Minor, Moderate, Major, and Catastrophic evenly spaced; default at 0 (Negligible)*]

*Accident Risks*

**10. Knowledge Deterioration**

A. What is the likelihood that in the next decade or two advanced AI technology leads to society's collective inability to determine what information is true or reliable? For example, this might happen if people over-rely on AI systems and blindly trust their outputs even if the output is wrong (e.g., fake news, generative AI like ChatGPT hallucinating or making things up).

[*Slider from 0% to 100%; labels every 20%, with 1% increments; default at 0*]

B. How impactful do you think a gradual distortion of knowledge might be for society?

[*100 pt Slider with labels Negligible, Minor, Moderate, Major, and Catastrophic evenly spaced; default at 0 (Negligible)*]

**11. Safety Systems Failures**

A. What is the likelihood that in the next decade or two an advanced AI system relied upon for public safety, such as disease detection, self-driving vehicles, or disaster warning systems, malfunctions and results in a significant loss of life?

[*Slider from 0% to 100%; labels every 20%, with 1% increments; default at 0*]

B. How impactful do you think the failure of safety critical advanced AI systems might be for society?

[*100 pt Slider with labels Negligible, Minor, Moderate, Major, and Catastrophic evenly spaced; default at 0 (Negligible)*]

*Geopolitical Risks*

**12. AI Arms Race**

A. What is the likelihood that in the next decade or two an advanced AI arms race between the United States and China leads to a direct military conflict between the two countries' armed forces?

[*Slider from 0% to 100%; labels every 20%, with 1% increments; default at 0*]

B. How impactful do you think an AI arms race inspired conflict between the US and China might be for society?

[*100 pt Slider with labels Negligible, Minor, Moderate, Major, and Catastrophic evenly spaced; default at 0 (Negligible)*]

**13. Global Power Shift**

A. What is the likelihood that in the next decade or two one or more countries develop a strategic advantage in AI giving rise to a global power imbalance that favors authoritarianism? For example, this might happen if one or more authoritarian governments create radically advanced AI before other countries or disproportionately to other countries, giving them some sort of strategic advantage and allowing them to enforce their values amongst the rest of the world.

[*Slider from 0% to 100%; labels every 20%, with 1% increments; default at 0*]

B. How impactful do you think losing an AI race might be for society?

[*100 pt Slider with labels Negligible, Minor, Moderate, Major, and Catastrophic evenly spaced; default at 0 (Negligible)*]

**14. Information Warfare**

A. What is the likelihood that in the next decade or two an adversary uses advanced AI to engage in information warfare to destabilize an adversary? For example, this might involve a sustained effort to provide very persuasive and convincing misinformation, like deep fakes, generated by advanced AI with the longer-term goal of destabilizing society, or, a similar use of persuasive advanced AI technology in order to influence the outcome of one or more federal elections.

[*Slider from 0% to 100%; labels every 20%, with 1% increments; default at 0*]

B. How impactful do you think a successful information warfare campaign by an adversary might be for affected members of society?

[*100 pt Slider with labels Negligible, Minor, Moderate, Major, and Catastrophic evenly spaced; default at 0 (Negligible)*]

*Environmental Risks*

**15. Energy Consumption**

A. What is the likelihood that in the next decade or two the computational resources required to support training, operation, and data storage for AI models consume enough energy that the resulting carbon emissions destabilize the global climate above and beyond the current median projections?

[*Slider from 0% to 100%; labels every 20%, with 1% increments; default at 0*]

B. How impactful do you think this risk that increased climate volatility might be for society?

[*100 pt Slider with labels Negligible, Minor, Moderate, Major, and Catastrophic evenly spaced; default at 0 (Negligible)*]

*Existential Risks*

**16. Dystopia**

A. What is the likelihood that in the next decade or two radically advanced AI overtakes existing global governments and implements a form of government that is not acceptable to the majority of humanity? For example, a society run by advanced AI might entail constant surveillance and no privacy, extreme poverty, a prohibition of religion or other critical liberties, or other restrictions causing widespread suffering or injustice.

[*Slider from 0% to 100%; labels every 20%, with 1% increments; default at 0*]

B. How impactful do you think an advanced AI dystopia might be for society?

[*100 pt Slider with labels Negligible, Minor, Moderate, Major, and Catastrophic evenly spaced; default at 0 (Negligible)*]

**17. Civilizational Collapse**

   A. What is the likelihood that in the next decade or two advanced AI will cause the collapse of human civilization resulting in global anarchy? For example, this might happen if AI automates systems like those for food production and distribution and we eventually lose control of our food supply resulting in mass famine; in this or other similar scenarios AI could lead to the deaths of so many people that governments cannot be sustained.

   [*Slider from 0% to 100%; labels every 20%, with 1% increments; default at 0*]

   B. How impactful do you think civilization collapse might be for society?

[*100 pt Slider with labels Negligible, Minor, Moderate, Major, and Catastrophic evenly spaced; default at 0 (Negligible)*]

**18. Extinction**

   A. What is the likelihood that in the next decade or two AI leads to human extinction?

   [*Slider from 0% to 100%; labels every 20%, with 1% increments; default at 0*]

   B. How impactful do you think human extinction might be for society?

[*100 pt Slider with labels Negligible, Minor, Moderate, Major, and Catastrophic evenly spaced; default at 0 (Negligible)*]

*Regulation*

19. In your opinion, what should we do about risks from advanced AI?
   A. We should accelerate the rate of AI progress.
   B. We should maintain the current rate of AI progress.
   C. We should slow down the rate of AI progress.
   D. We should take an indefinite pause on AI progress.

20. In your opinion, who should manage risks from advanced AI?
   A. The companies developing AI technologies should manage the majority of risks.
   B. National governments should manage the majority of risks.
   C. International treaties, intergovernmental organizations, and non-governmental organizations should manage the majority of risks.

*Demographic Questions*

Do you consider yourself an AI expert?
- Definitely not
- Probably not
- Might or might not
- Probably yes
- Definitely yes

What is your profession?
- Text box

What is the highest level of education you have completed?
- Some high school
- Graduated from high school (Diploma/GED or equivalent)
- Some college, no degree
- Completed associate's degree
- Completed bachelor's degree
- Completed master's degree
- Completed professional degree beyond a bachelor's degree (e.g., M.D., J.D.)
- Completed doctorate degree

On what continent(s) did you attend high school?
- Asia
- Africa
- Europe
- North America
- Oceania
- South America

What is your household income per year? (please provide your pre-tax income)
- Under $20,000
- Between $20,000 and $49,999
- Between $50,000 and $79,999
- Between $80,000 and $99,999
- Between $100,000 and $149,999
- $150,000 or more

Are you registered to vote in US elections?
- No, I am NOT registered to vote
- Yes, I am registered to vote

Generally speaking, do you think of yourself as a ...?
- Republican
- Democrat
- Independent
- Other
- Not sure

What is your gender?
- Man
- Woman

- Non-binary or other gender identification
- Prefer not to say

Do you approve or disapprove of the way Joe Biden is handling his job as President?
- Strongly approve
- Somewhat approve
- Somewhat disapprove
- Strongly disapprove
- Don't know / No opinion

What is your current age in years?
- Text box

Which of the following best matches your racial identity?
- American Indian or Alaska Native
- Asian or Asian American
- Black or African American
- Native Hawaiian or Pacific Islander
- White
- Other race
- Identify with two or more races

Do you consider yourself to be of Hispanic, Latino, or Spanish origin?
- Yes
- No, I am NOT of Hispanic, Latino, or Spanish origin

In which US State or District do you currently live?
- Dropdown list

How would you describe the place where you live?
- City
- Suburb
- Town
- Rural area
- Other

Would you describe yourself as a "born-again" or evangelical Christian, or not?
- Yes
- No

## Materials B: Expert Recruitment Email

**Subject:** Survey on Societal-Scale AI Risks

Hi,

You are invited to participate in a survey on societal-scale AI risks. The goal of this research is to better understand experts' perceptions of different types of societal-scale risks from AI.

We are seeking expert researchers with expertise on some dimension of AI, and you have been invited given your previous research related to AI.

The survey will take approximately 10 minutes to complete. If you complete the survey, you will be compensated for your time with a $50 gift card. Please be aware that the survey will close once we have recruited a fixed number of respondents.

If you have any questions about your potential participation in the study, please contact Dr. Ross Gruetzemacher at 1845 Fairmount St., Wichita, KS 67226; telephone +1 (316) 979-6242; email ross.gruetzemacher@wichita.edu.

If you are interested in participating, please click here to review the consent form and begin the survey. Thank you for your consideration.

Best,

Ross Gruetzemacher, PhD
Assistant Professor, Wichita State University
Executive Director, Transformative Futures Institute
Research Affiliate, Centre for the Study of Existential Risk
Fellow, Foresight Institute

## Materials C: Voter Recruitment

*Advertisement*

Title: Survey About Societal Issues
Summary: A study to understand US residents' opinions on the likelihood and impact of various societal-scale risks from artificial intelligence (AI). If you decide to participate, you will be asked to complete a survey that will take 15 minutes.

*Consent*

I am Ross Gruetzemacher, an Assistant Professor of Business Analytics at Wichita State University, and I am contacting you because you are a US resident. I am recruiting research participants to help with a study to understand US residents' opinions on the likelihood and impact of various societal-scale risks from AI. If you decide to participate, you will be asked to complete a survey that will take 15-20 minutes.

In addition to the survey questions, we will request age, gender, educational history, country of high school, knowledge of AI, income, US voter registration status, political party, presidential approval, race, state of residence, population density, and religion.

There are no personal benefits or anticipated risks to participating in this study. However, if you feel uncomfortable with a question, you may skip it. Participation is voluntary, and you can stop taking the survey at any time.

You will receive $10 for your participation.

We will work to make sure that no one sees your survey responses without approval. While the survey is active your data will be stored with Qualtrics, and after the completion of the survey your data will be stored on an encrypted drive in a computer on the Wichita State University campus. Email addresses will be necessary for providing you with a gift certificate upon completion, but this is the only purpose of your email address, and it will not be retained with the data stored at Wichita State University. Because we are using the Internet, there is a chance that someone could access your online responses without permission. In some cases, this information could be used to identify you.

If you have any questions, please contact ross.gruetzemacher@wichita.edu. For questions about the rights of research participants, you may contact the Office of Research at Wichita State University, 1845 Fairmount Street, Wichita, KS 67260-0007, and telephone (316) 978-3285.

You are under no obligation to participate in this study. By selecting "Yes" below, you are indicating that:
• You have read (or someone has read to you) the information provided above,
• You are aware that this is a research study,
• You have voluntarily decided to participate.

If you would like a copy of the Consent Form, you can download it by clicking here.
*
I have read the above and agree to participate in this survey.
Yes
No
*
I am age 18 or over.
Yes
No